\documentclass[natbib]{svjour3}
\smartqed

\usepackage{graphicx}

\usepackage{aps-bibstyle}  
\usepackage{graphicx}
\usepackage{dcolumn}
\usepackage{bm}
\usepackage{hyperref}

\def\CQG{{\it Class. Quantum Gravity} }

\def\IJMP{{\it Int. J. Mod. Phys.} }

\def\NP{{\it Nucl. Phys.} }

\def\PRL{{\it Phys. Rev. Lett.} }

\def\om{\omega}  \def\De{\Delta}

 \def\frac#1#2{{\textstyle{{#1}\over
{#2}}}} 
\def\lsim{\mathrel{\rlap{\lower4pt\hbox{\hskip1pt$\sim$}}
\raise1pt\hbox{$<$}}}
\def\gsim{\mathrel{\rlap{\lower4pt\hbox{\hskip1pt$\sim$}}
\raise1pt\hbox{$>$}}} \def\sqr#1#2{{\vcenter{\vbox{\hrule height.#2pt
\hbox{\vrule width.#2pt height#1pt \kern#1pt \vrule width.#2pt} \hrule
height.#2pt}}}}

\def\beq{\begin{equation}} \def\eeq{\end{equation}}
\def\beqa{\begin{eqnarray}} \def\eeqa{\end{eqnarray}}
\def\eq#1{Eq. (\ref{#1})}

\begin{document}

\title{Probing the Flyby Anomaly with the future STE-QUEST mission\thanks{J. P\'aramos acknowledges support of the Funda\c c\~ao para a Ci\^encia e Tecnologia under the project PDTC/FIS/111362/2009. G. Hechenblaikner acknowledges support from the European Space Agency under contract number 4000105368/12/NL/HB for the ongoing STE-QUEST mission assessment activities, some results of which enter the analyses of this paper.}}

\author{Jorge P\'aramos \and Gerald Hechenblaikner}
\institute{J. P\'aramos \at Instituto de Plasmas e Fus\~ao Nuclear, Instituto Superior T\'ecnico, Universidade T\'ecnica de Lisboa, Av. Rovisco Pais 1, 1049-001 Lisboa, Portugal\\\email{paramos@ist.edu} \and G. Hechenblaikner \at Astrium Satellites GmbH, 88039 Friedrichshafen, Germany\\\email{Gerald.Hechenblaikner (at) astrium.eads.net}}

\date{\today}

\maketitle

\begin{abstract}
In this study, we demonstrate that the flyby anomaly, an unexpected acceleration detected in some of the gravitational assists of the Galileo, NEAR, Cassini and Rosetta spacecraft, could be probed by accurate orbital tracking available in the proposed Space-Time Explorer and Quantum Equivalence Principle Space Test (STE-QUEST); following a recent work, we focus on the similarity between an hyperbolic flyby and the perigee passage in a highly elliptic orbit of the latter, as well as its Global Navigation Satellite System precise orbital determination capabilities.

\keywords{Flyby Anomaly \and GNSS \and STE-QUEST mission}
\PACS{95.10.Ce,07.87.+v,95.30.Sf}

\end{abstract}

\section{The Flyby Anomaly}

Since the end of the last century, an analysis of the Earth gravitational assist maneuvers of the Galileo, NEAR, Cassini and Rosetta spacecraft has disclosed an anomalous  velocity change after some of these flybys [\cite{Anderson2008}]. Subsequent flybys of the Galileo and Rosetta missions were met with some expectation, in hope of reproducing this phenomenon. As Table~\ref{flyby_table} [\cite{Anderson2008,Antreasian1998,AdlerUpdate}] shows, these events yielded no further evidence of such a flyby anomaly --- in the case of the second Galileo flyby, this occurs because of the high uncertainty in the determination of the atmospheric drag, which is enhanced due to the much lower perigee of $\sim 300~km$. 
 
Excluding Cassini, none of the spacecraft had any available Deep Space Network (DSN) tracking during perigee (an approximate four hour gap), and during the remaining period (covered by DSN availability) the interval of approx. 10 s between data points was too coarse for an accurate characterization of the effect. For this reason, no acceleration profile exists for the  crucial perigee passages, so that the flyby anomaly cannot be characterized as an additional force acting upon the bodies. Instead, the flyby anomaly is revealed by the inability to trace a single hyperbolic arc ({\it i.e.} an open orbit) to the whole maneuver: instead, two distinct ``incoming'' and ``outgoing'' arcs were fitted to the spacecraft trajectory, with the slight difference between them being due to an additional boost $\Delta v$ at perigee (see Table \ref{flyby_table}). The latter can be regarded as the currently available observable signaling the anomalous events.

Thus, one can only assign an averaged value to the putative force causing such deviations from the expected path of the spacecraft: this is found to be of the order of $10^{-4}~{\rm m/s^2}$ [\cite{Antreasian1998}]. Albeit tentative, this enables the direct comparison with several known sources for perturbations to the hyperbolic trajectories, {\it e.g.} Earth oblateness, other Solar System bodies, relativistic corrections, atmospheric drag, Earth albedo and infrared emissions, ocean or solid tides, solar pressure,  spacecraft charging, magnetic moments, solar wind and spin-rotation coupling, {\it etc.} [\cite{Antreasian1998,Lammerzahl2006}]. 

A list of the magnitudes of all relevant effects is given in Table \ref{error_sources_table} [\cite{Antreasian1998,Lammerzahl2006}]: clearly, these are all orders of magnitude smaller than the required value, with the exception of Earth oblateness. This might hint that possible errors in the gravitational model of the Earth could be the origin of the flyby anomaly. Notice, however, that measurements throughout the years have produced an ever-shrinking scope of allowed values for the latter, with each new result confirming the previously available (larger) range. This ``zooming'' in on $J_2 \approx -1.082626683 \times 10^{-3}$ hints that no improvements on its determination should deviate from the currently considered constraints; since any putative changes to the latter compatible with a flyby anomaly fall outside of this interval, such an explanation is deemed implausible [\cite{Antreasian1998}]g.

As a result, the yet unknown origin of the flyby anomaly could signal the presence of new or "exotic" physics at play, a possibility which should not be taken lightly: indeed, while a new force could perhaps account for the flyby anomaly, it should also modify a plethora of other phenomena, from planetary orbits to E\"otv\"os-type experiments. Furthermore, no clear cut fundamental motivation exists for such a short ranged force (see Refs. \cite{Lammerzahl2006} and \cite{paramos2012} for a brief overview of some proposed physical mechanisms).

Amongst the growing number of proposals, one highlights the empirical formula proposed in Ref. \cite{Anderson2008}  to fit the flyby relative velocity change as a function of the declinations of the incoming and outgoing asymptotic velocity vectors, $\delta_i$ and $\delta_o $, respectively

\begin{equation}
	{\Delta V_\infty \over V_\infty} = K (\cos \delta_i - \cos \delta_o), \label{modelPRL}
\end{equation}
\noindent where the constant $K$ is expressed in terms of the Earth's rotation velocity $\omega_E$, its radius $R_E$ and the speed of light $c$ as $K = 2\om_e R_e/c$. This identification is reminiscent of the term found in the outer metric due to a rotating body [\cite{Ashby}],

\begin{equation}
	ds^2 = \left(1 + 2{V - \Phi_0 \over c^2} \right)(c~dt)^2 - \left(1 - 2{V \over c^2} \right)(dr^2+r^2 d\Omega^2),
\end{equation}
with
\begin{equation}
	{\Phi_0 \over c^2} = {V_0 \over c^2} - {1 \over 2} \left( \omega_e R_e \over c \right)^2,
\end{equation}
where $d\Omega^2 = d\theta^2 + \sin^2\theta d\phi^2$, $V_0 $ is the value of the Newtonian potential $V(r)$ at the equator, $\omega_e$ the Earth's rotational velocity and $R_e$ its radius.

However, such a suggestive relation is misleading: \eq{modelPRL} is impossible to derive from General Relativity (GR), and the flyby anomaly is much higher than the relativistic effects induced by the rotation of the Earth: the geodetic effect and frame dragging.

One could argue that the flyby anomaly only affects hyperbolic orbits, thus explaining why it hasn't been detected in satellites in low Earth orbit --- namely the Gravity Probe B mission [\cite{GPB}] which, at a altitude of $\sim 600~{\rm km}$, accurately measured both effects but did not detect an anomalous acceleration with magnitude comparable to the $10^{-4}~{\rm m/s^2}$ scale of the flyby anomaly, although it travels well inside the assumed zone where the latter is present. This possibility would imply an explicit breaking of the Equivalence Principle, according to which the acceleration of a body is independent of its properties (namely, its energy) and (in the absence of non-gravitational forces) only reflects the gravitational field [\cite{review}].
 
Furthermore, \eq{modelPRL} predicted that the subsequent two flybys by the Rosetta probe (in 2007 and 2008) should experience an anomalous increase in $V_\infty$ of respectively $0.98$ and $1.09$ mm/s [\cite{Busack}] --- while the subsequent analysis of the tracking data was consistent with no flyby anomaly being present.

With the above in mind, a more sober explanation for the flyby anomaly should not be dismissed: some poorly modelled behaviour of the affected spacecrafts could be the culprit, and also explain why the anomalous $\Delta v$ varies so widely with the different designs and gravitational assists.

\begin{table}
\begin{center}
\caption{\label{flyby_table}Summary of orbital parameters of the considered Earth flybys.}
\begin{tabular}{@{}ccccccc}
					\hline\hline

Mission	& Date	& $e$	& Perigee		& $v_\infty$		& $\Delta v_\infty$	& $\Delta v_\infty / v_\infty$ \\
					&		&		& $({\rm km})$	& $({\rm km/s})$ 	& $({\rm mm/s})$ 	& $(10^{-6})$ \\
					\hline
			Galileo	& 1990	& $2.47$& $959.9$		& $8.949$			& $3.92 \pm 0.08$	& $0.438$  \\  
			Galileo	& 1992	& $3.32$& $303.1$		& $8.877$			& $\sim 0 $		& $-0.518$ \\  
			NEAR	& 1998	& $1.81$& $538.8$		& $6.851$			& $13.46 \pm 0.13$	& $1.96$   \\
			Cassini	& 1999	& $5.8$	& $1173$		& $16.01$			& $-2 \pm 1$		& $-0.125$ \\  
			Rosetta	& 2005	& $1.327$& $1954$		& $3.863$			& $1.80 \pm 0.05$	& $0.466$  \\
			MESSENGER& 2005	& 1.360		& $2347$		& $4.056$			& $0.02 \pm 0.01$	& $0.0049$ \\
			Rosetta	& 2007	& 3.562		& $ 5322 $	& 9.36					& $\sim 0$			& - \\
			Rosetta	& 2009	& 2.956		& $2483$		& 9.38				& $\sim 0$			& - \\

\hline\hline

\end{tabular}
\end{center}
\end{table}


\begin{table}
\begin{center}
\caption{\label{error_sources_table}List of orders of magnitude of possible error sources during Earth flybys.}
\begin{tabular}{@{}cc}

\hline\hline
Effect				& Order of Magnitude \\
								& $({\rm m/s^2})$ \\
		\hline											
			Earth oblateness		& $10^{-2}$	\\
			Other Solar System bodies & $10^{-5}$	\\
			Relativistic effects		& $10^{-7}$	\\
			Atmospheric drag	& $10^{-7}$	\\
			Ocean and Earth tides	& $10^{-7}$	\\
			Solar pressure		& $10^{-7}$	\\
			Earth infrared		& $10^{-7}$	\\
			Spacecraft charge	& $10^{-8}$	\\
			Earth albedo		& $10^{-9}$	\\
			Solar wind			& $10^{-9}$	\\
			Magnetic moment		& $10^{-15}$	\\

\hline\hline

\end{tabular}
\end{center}
\end{table}

As discussed above, one of the major caveats in the study of the flyby anomaly is its lack of spatial resolution within the crucial perigee passage, which makes it impossible to directly compute an acceleration or, more rigorously, a post-processing treatment of the spacecraft's path through an orbital determination program that dynamically simulates all the relevant interactions.

To be able to distinguish between a true anomaly or a possible misreading due to a corrupted Doppler signal (in itself a very interesting prospect), the idea of using the Global Navigation Satellite System (GNSS) to determine the position and velocity was put forward in Ref. [\cite{paramos2012}]. The former distinction is possible as GNSS is able not only to measure frequency shifts, but relies also on the propagation time to derive the position and velocity of receivers.

This proposal also argued that, in order to be able to reproduce the flybys, and thus greatly increase the statistical significance of the ensuing results, an hyperbolic orbit could be abandoned for a highly elliptical one, as long as the perigee altitude and velocity are of the same order of magnitude [\cite{paramos2012}]. As it turns out, a mission with the required characteristics is being planned, providing an excellent test bed for this enduring puzzle in physics: the Space-Time Explorer and Quantum Equivalence Principle Space Test (STE-QUEST) mission proposal for a medium mission for ESA uses GNSS tracking (complemented with laser ranging, if required) and is designed to travel along a highly elliptic orbit.

In the next section we give a brief overview of the general features of STE-QUEST, including its science objectives, precise orbit determination capabilities, orbital motion and perturbative effects. Following this, the case for the use of STE-QUEST in measuring the flyby anomaly is presented, with an emphasis on its behaviour in the crucial perigee passages and the strategy deployed to assess the presence of an anomalous force.

\section{The STE-QUEST mission proposal}
\subsection{Mission objectives}
The Space-Time Explorer and Quantum Equivalence Space Test (STE-QUEST) [\cite{Sch10,SciRD}] is a medium-size mission candidate for launch in 2022/2024. It was selected by the European Space Agency (ESA), together with three other candidates, for an assessment study until mid-2013. This started with an ESA-internal assessment,  followed by (currently ongoing) mission assessment studies performed by competitive industrial teams and instrument definition studies performed by the scientific collaborators. The mission aims to explore the realm of gravity, the arguably least understood and only fundamental force persistently defying attempts towards unification with the weak, strong, and electromagnetic interactions in an ultimate theory of everything. Whereas Einstein's metric theory of General Relativity (GR) describes the effects of gravity on macroscopic scales, it cannot be easily reconciled with Quantum Physics --- which, on the other hand, has proved extremely successful in describing all other fundamental forces and constitutes the foundation of modern physics. It would therefore not be unexpected that GR breaks down at some point and proves to be insufficient in predicting the outcome of measurements of an increasingly higher level of accuracy. The accuracy of standard laboratory experiments and astronomical observations has so far been insufficient to reach the limits of GR and detect a deviation from its predictions: STE-QUEST aims to push the frontiers forward in measurement and verification of the three cornerstones of Einstein's Equivalence Principle (EEP), which describes the properties of space-time and matter:
\begin{itemize}
    \item Local Position Invariance (LPI): all non-gravitational experiments performed in locally free falling frames will yield the same results independent of the velocity of the reference frame;
    \item Local Lorentz Invariance (LLI): the outcome of any non-gravitational experiment is independent of the time and place where it is performed;
    \item Weak Equivalence Principle (WEP): the trajectories of freely falling test bodies are independent of their structure and composition. In a simplified Newtonian picture this means that every test-mass must fall with the same acceleration in a given gravitational field.
\end{itemize}

During its five year mission, the STE-QUEST mission has three primary science objectives, aimed at probing the three cornerstones of EEP, as listed below:

\begin{itemize}
\item Gravitational Redshift tests
	\begin{description}
	\item[1:] Earth gravitational red-shift --- measure to a fractional frequency uncertainty better than $2\times 10^{-7}$ (goal $4\times 10^{-8}$)
	\item[2:] Sun gravitational red-shift --- measure to a fractional frequency uncertainty better than $2\times 10^{-6}$ (goal $6\times 10^{-7}$)
	\end{description}
\item Weak Equivalence Principle Test
	\begin{description}
	\item[3:] Universality of the free propagation of matter  waves --- test to an uncertainty in the E\"otv\"os parameter better than $\eta=1\times 10^{-15}$
	\end{description}
\end{itemize}

\noindent To achieve these objectives two instruments and some additional support equipment are used onboard the spacecraft.

\subsection{Measurement Strategy}

The first instrument, an atomic clock, is used for frequency comparison against ground clocks (located close to the sites of three ground terminals) to determine gravitational red-shifts in the gravitational field of the Earth or the Sun, respectively, in search of possible violations of gravitational time dilation and Local Position Invariance. Time and frequency data are exchanged between space- and ground-clocks via two-way microwave and optical links. In that respect, it should be pointed out that the soon to be launched ACES (Atomic Clock Ensemble in Space) mission aims to perform measurements between ground- and space-clocks using micro-wave clocks in a similar fashion --- albeit at lower target accuracy than STE-QUEST and using the International Space Station (ISS) as the platform for the space-clock PHARAO [\cite{ACES}].

 The second instrument, an atom interferometer, is used to  measure the differential acceleration between two atomic species  under free fall conditions, in an attempt to detect possible violations of the universality of free fall in the quantum regime. Other proposals to test EEP include {\it e.g.} the Galileo Galilei dedicated spacecraft [\cite{GG}] and the LATOR laser ranging experiment between two spacecraft and the ISS [\cite{LATOR}].

In order to achieve the high measurement performance goals required in Mission Objectives 1 and 2, it is necessary to have an accurate space-clock combined with even more accurate ground clocks, described by fractional frequency inaccuracies on the order of $10^{-16}$ and $10^{-18}$, respectively. Similarly, the fractional frequency instabilities for both clocks and microwave links set ambitious targets which can only be achieved after long integration times, on the order of $10^{5}$ s or longer. However, in addition to these requirements for clocks and links, the provision of highly accurate Precise Orbit Determination (POD) equipment is another crucial factor for the overall measurement performance.

On the one hand, it is important to track the position of the space-clock in relation to a ground clock to determine the exact difference in gravitational potential and hence the predicted red-shift between the two clocks; on the other hand, it is also required to know the respective velocities to calculate and consider the relativistic Doppler effect in the post-processing and data analysis. For this reason, a POD accuracy (in post-processing) of up to 0.5 m in position and of up to 0.2 mm/s in velocity is required so that POD error contributions do not dominate the measurement performance. Investigations are currently being performed on how this could be accomplished through a combination of a precision multi-GNSS receiver together with Corner Cube Reflectors for laser ranging.

\subsection{The reference orbit}
\begin{figure}
 \includegraphics[width=0.95 \linewidth]{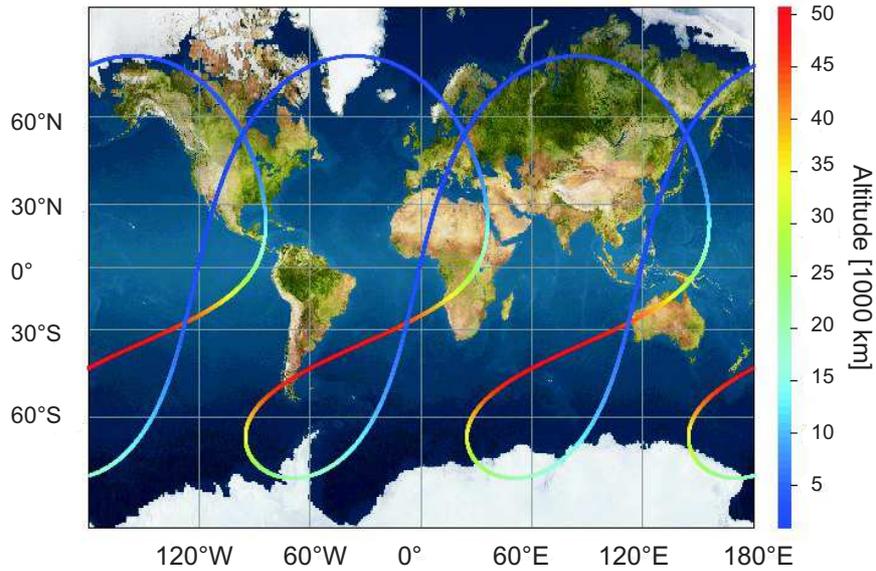}%
 \caption{\label{fig_1}Ground track of the STE-QUEST 16 h orbit with a 3:2 resonance.}
 \end{figure}

Atomic clock measurements require a high-eccentric-orbit (HEO) for maximal difference in gravitational potential between perigee and apogee, combined with long contact periods with the ground terminals for optimal measurement performance. In parallel, atom interferometer measurements require large gravity gradients to maximize the E\"otv\"os parameter and are therefore best performed close to Earth --- ideally at an altitude where the perturbation through air drag is sufficiently small.
\\The STE-QUEST baseline orbit, designed to optimize and balance the measurement requirements for both instruments, is described in more detail in Ref. \cite{Mag}. Its Kepler elements are listed in table \ref{table2}. The orbit period of 16 h is chosen such that the corresponding 3:2 resonance ground track features perigee locations in the vicinity of the three ground terminals (located at Boulder, Torino, and Tokyo) during the initial phase of the mission (depicted in Fig. \ref{fig_1}).
Due to the non-spherical character of the Earth gravitational field ({\it i.e.} a non-vanishing $J_2$ term), there is a rotation of the inclination plane. This also affects a drift of the right ascension of the ascending node (RAAN), which can be adjusted through slight changes of the orbital period and therefore of the semi-major axis. Another effect of the non-spherical gravitational potential is a rotation of the line of apses and a corresponding drift of the argument or perigee on the order of 1/20 deg. per day, which gradually moves it from North to South (and {\it vice versa} for the apogee) during the five years of mission. For this reason, contact with the Northern ground terminals at perigee is only possible during the initial phase of the mission, whilst at later stages --- when apogee has drifted North --- good common view contacts with several ground terminals (required for objective II) are ensured during apogee passage.

\begin{table}
\begin{center}
\caption{\label{table2}The STE-QUEST orbit parameters.}
\begin{tabular}{@{}ll}

\hline\hline
Epoch&01 June  2022\\
&21:36:00 UTC\\
\hline
Semi-major Axis (km)& 32025\\
Eccentricity & 0.773\\
Inclination ($^{\circ}$)& 72.071\\
Argument of Perigee ($^{\circ}$)& 43.546\\
RAAN ($^{\circ}$)&89.132\\
True Anomaly ($^{\circ}$)& 28.648\\
Average Drift in argument of perigee ($^{\circ}$/d)& -0.055\\
Average Drift in RAAN $^{\circ}$/d)&-0.063\\

orbital period&16 h\\
apogee altitude & 50 395 km\\
perigee altitude (initial) & 899 km \\
\hline\hline

\end{tabular}
\end{center}
\end{table}

\subsection{Variable flyby altitudes and spacecraft pointing}
\begin{figure}
 \includegraphics[width=0.98 \linewidth]{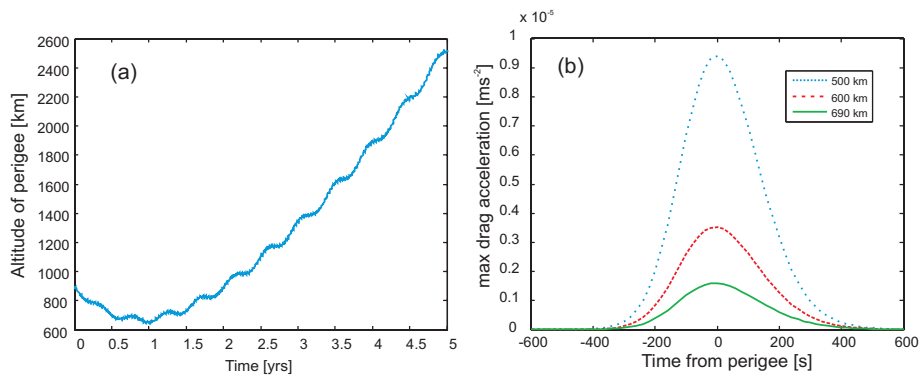}%
 \caption{\label{fig_2}(a) The perigee altitude of the STE-QUEST orbit is plotted against the mission time. (b) Drag forces encountered for the orbit are plotted for perigee altitudes of 500 km (blue, dotted), 600 km (red, dashed), and 690 km (green, solid).}
 \end{figure}

A very important feature of the STE-QUEST orbit is the variation of perigee altitude due to third body effects from the Sun and the Moon throughout the mission; a plot of the variation of orbital altitude is given in Fig. \ref{fig_2}a. The perigee altitude reduces from initially 900 km to approximately 690 km after one year, before increasing to more than 2000 km after 4 years. Comparing to table \ref{flyby_table} of the previous section, it becomes apparent that the interesting range of altitudes where anomalies have previously been observed (500 km to 2000 km) is almost fully covered during the first four years of the STE-QUEST mission. It is important to note that the STE-QUEST orbit has been designed with the air drag accelerations experienced during perigee passage in mind: these are typically required to be below a level of approximately $10^{-6}~{\rm m/s^2}$, so that the instrument performance of the atom interferometer is not compromised. Figure \ref{fig_2}b plots the drag accelerations experienced during perigee passage for three different perigee altitudes, assuming high solar activity and a maximal cross-section of $23~m^2$ for the spacecraft and solar arrays. It is only for the minimum altitude of 690 km that the drag acceleration slightly exceeds the requirement for these rather conservative assumptions, whereas for higher altitudes the drag forces are unproblematic. Nevertheless, considerable uncertainties on the exact magnitude of the drag forces and the associated torques make it desirable to compensate the drag though use of dedicated micro-propulsion system (MPS). The same system is also used for attitude control instead of reaction wheels, so that problems relating to micro-vibrations arising from the latter can be avoided. A final decision on the inclusion of the MPS and whether this would be open-loop or closed-loop has not been made, and is deferred to the next phase of the mission assessment.

In order to avoid undesirable rotations of the spacecraft during sensitive atom interferometer measurements, which are generally restricted to altitudes below 3000 km where gravity gradients are sufficiently large, the spacecraft is kept inertial during the perigee passage of approximately 1900 s. As was pointed out in previous discussions of the flyby anomaly, this easily covers the time window of a few minutes during which the flyby anomaly is expected to occur. No thruster operation would therefore disguise a potential anomaly; even if the MPS was operated to compensate the mean drag force, the typical thrust levels are only in the range of $10^{-6}~{\rm m/s^2}$ and therefore very small compared to the expected anomalous acceleration, of the order of $10^{-4}~{\rm m/s^2}$. Furthermore, the thrust levels are continuously recorded, and that data could be used in any post-encounter analysis of the flyby. In table \ref{table3} we compare the parameters of the STE-QUEST orbit to to those of similar parabolic and hyperbolic orbits with the same perigee altitude.


\begin{table}
\begin{center}
\caption{\label{table3}Comparison between STE-QUEST orbit and equivalent parabolic and hyperbolic orbits.}
\begin{tabular}{@{}llll}

\hline\hline
Orbital parameter&HE Orbit&Parabolic&Hyperbolic\\
\hline
Perigee Altitude&690 km&690 km&690 km\\
Apogee Altitude&50395 km&-&-\\
Velocity at Perigee&10.01 km/s&10.62 km/s&13.00 km/s\\
Eccentricity&0.78&1.00&2.00\\
Orbital Period&16 h&-&-\\
\hline\hline

\end{tabular}
\end{center}
\end{table}


\section{Measuring the flyby anomaly with STE-QUEST}
\subsection{Precise Orbit Determination}
As mentioned in previous sections, a key ingredient in a successful characterization of the flyby anomaly is the accurate determination of the position and velocity of the spacecraft during gravitational assist. As a reminder, the typical momentum transferred during flyby is on the order of several mm/s, so that the specified tracking accuracy of 0.2 mm/s for STE-QUEST is easily sufficient to achieve this goal. The likely inclusion of an additional accelerometer on-board the spacecraft (as a means of implementing closed-loop drag compensation) would provide another means of improving measurement accuracy and isolating external perturbations from the recorded data on spacecraft position and momentum: with a projected sensibility in the $10^{-8}  - 10^{-4}~{\rm m/s^2}$ range, it would provide a direct measurement of non-gravitational forces affecting the spacecraft, thus helping to establish if the putative flyby anomaly is of gravitational origin or not.

Furthermore, the covered range of perigee altitudes throughout the mission duration (ranging from 690 km to 2000 km) ensures not only that the STE-QUEST spacecraft will probe the regions where the flyby anomaly was reported, but also that it will be able to profile the variation of its magnitude with varying distance to the Earth --- something that cannot be directly extracted from the available gravitational assists depicted in Table \ref{flyby_table}, as the different altitudes were probed with spacecrafts with distinct designs and features that could distort the impact of the anomaly.

The STE-QUEST mission provides yet another advantageous feature for testing the origin of the flyby anomaly: its attitude control system. Although the planned mission assumes a constant attitude throughout perigee passages (as required by its established scientific objectives), some variations of its orientation can be easily accommodated during its mission lifetime: a variation of the magnitude and direction of the anomaly  with the attitude would hint that it is due to unaccounted effects within the spacecraft itself ({\it e.g.} outgassing), not a deviation from the known law of gravity. This possibility is enabled because the MPS used for attitude control is accurately modelled and its typical acceleration level is two orders of magnitude below the $10^{-4}~{\rm m/s^2}$ figure of merit for the flyby anomaly, as discussed before.

\subsection{External perturbations}

There are a number of perturbations acting on the spacecraft which lead to momentum being transferred to the spacecraft and deviations from a simple osculating orbit. These perturbative forces are summarized in table \ref{table4}, having assumed a spacecraft mass of $m_s=2000$ kg and a perigee passage time (defined as the time below 3000 km altitude) of approximately 1900 s --- during which the spacecraft remains inertial and covers an angle of 64 degrees around perigee.

As can be seen, only the effects of the Earth oblateness, lunar attraction and ocean tides compare with the reported typical values of the flyby anomaly. Fortunately, these forces are very deterministic and accurately reflected in every reasonable orbit propagator, so that they can be ruled out as origin of the observed anomalies. One may also recall that the first has been shown not to be the cause for the anomalous increase in velocity (as discussed in a preceding section), while the lunar and ocean tides effects have a clear temporal variation than allows for its discrimination throughout consecutive flybys.

\begin{table}

\begin{center}
\caption{\label{table4}Perturbations acting on the spacecraft and estimated momentum gain.}
\begin{tabular}{@{}llll}

\hline\hline
Effect& Acceleration& $|\De v|$ per passage\\
&(m/s$^2$)& (mm/s)\\
\hline
Earth oblateness&$< 2.1 \times 10^{-2}$& $< 2.1 \times 10^4$ \\
Third Body (Moon)&$<1.5 \times 10^{-6}$&$<1.2$\\
Ocean Tides&$<10^{-5} $&$<19 $\\
Relativistic Effects &$< 1.27 \times 10^{-8}$& $ \sim 0 $ \\
Atmospheric Drag&$<1.6\times 10^{-6}$&$<0.54$\\
Solar radiation pressure&$<1.2\times 10^{-7}$&$<0.23$\\
Earth albedo&$<1.2\times 10^{-8}$&$<0.02$\\
Solar wind&$<5.0\times 10^{-8}$&$<0.1$\\
\hline\hline

\end{tabular}
\end{center}
\end{table}

The drag forces experienced by the STE-QUEST spacecraft under worst case assumptions during perigee passage are given by the green line in Fig. \ref{fig_2}b. The total transferred (negative) momentum per orbit is approximately 0.5 mm/s, considerably less than typical values of several mm/s encountered in previously recorded flyby anomalies (see Table \ref{flyby_table}); under less conservative conditions, the momentum transfer is more than an order of magnitude lower than this. Some other perturbations such as solar radiation pressure, Earth albedo and solar wind, although difficult to predict accurately, are shown to be negligible.

\section{Conclusions}

This study shows that the flyby anomaly can be tested with the STE-QUEST mission, since its precise orbital determination accuracy is below the magnitude $10^{-4}~{\rm m/s^2}$ characterizing the former, and other perturbations are either below this level or can be successfully modeled and accounted for. The selected orbit for STE-QUEST is highly elliptical, and its perigee reaches an altitude as low as $690~{\rm km}$, enabling consecutive passages that allow for a replication of the typical velocities and altitudes of the reported flyby anomaly. The temporal resolution of the proposed GNSS tracking (possibly complemented with laser ranging) allows for the detection of any unaccounted acceleration behind the flyby anomaly, greatly improving upon the current situation where the latter is signaled by the mismatch between incoming and outgoing hyperbolic arcs.

This additional scientific application can be achieved by the STE-QUEST mission at no extra cost in hardware and with no additional operational maneuvers; it does so by utilizing the provided onboard equipment and exploiting the baselined operational scenario, whilst relying on a natural evolution of the uncontrolled baseline orbit. Furthermore, if a reorientation using its MPS is considered for a small number of perigee passages, a possible dependence of the flyby anomaly on the attitude and cross section of the spacecraft could help shed some light into the origin of this puzzling issue in contemporary physics.

\begin{acknowledgements}
The authors would like to thank O. Bertolami (IPFN \& Univ. Porto), F. Francisco (IPFN), P. J. S. Gil (IST \& IDMEC) and U. Johann (EADS Astrium) for fruitful and elucidating discussions, S. Sch\"aff (ASTOS solutions) for his contributions to the mission analysis, and M. Vitelli as well as J. Loehr (EADS Astrium) for their contributions to the drag force analysis.
\end{acknowledgements}


\begin{thebibliography}{}

\bibitem[\protect\citeauthoryear{Anderson {\it et al.} }{2008}]{Anderson2008}
J. D. Anderson, {\it et al.}, \PRL {\bf 100}, 091102 (2008).

\bibitem[\protect\citeauthoryear{Antreasian and Guinn}{1998}]{Antreasian1998}
P. G. Antreasian and J. R. Guinn, in \textit{AIAA/AAS Astrodynamics Specialist Conference and Exhibit}, AIAA Paper No 98-4287 (1998).

\bibitem[\protect\citeauthoryear{Adler}{2011}]{AdlerUpdate}
S. L. Adler, astro-ph.EP/1112.5426.

\bibitem[\protect\citeauthoryear{Laemmerzahl, Preuss and Dittus}{2006}]{Lammerzahl2006}
C. Laemmerzahl, O. Preuss and H. Dittus, in \textit{Lasers, Clocks, and Drag-Free: Technologies for Future Exploration in Space and Tests of Gravity} (Springer Verlag, 2006), gr-qc/0604052.

\bibitem[\protect\citeauthoryear{Bertolami {\it et al.} }{2012}]{paramos2012}
O. Bertolami, F. Francisco, P. J. S. Gil and J. P\'aramos, \IJMP {\bf D 21}, 1250035 (2012).

\bibitem[\protect\citeauthoryear{Ashby}{2003}]{Ashby}
N. Ashby, {\it Liv. Rev. Rel.} {\bf 6}, 1 (2003).

\bibitem[\protect\citeauthoryear{Everitt {\it et al.} }{2011}]{GPB}
C. W. F. Everitt {\it et al.}, \PRL {\bf 106}, 221101 (2011).

\bibitem[\protect\citeauthoryear{Bertolami, P\'aramos and Turyshev}{2006}]{review}
O. Bertolami, J. P\'aramos and S. G. Turyshev, in \textit{Lasers, Clocks, and Drag-Free: Technologies for Future Exploration in Space and Tests of Gravity} (Springer Verlag, 2006), gr-qc/0602016.

\bibitem[\protect\citeauthoryear{Busack}{2010}]{Busack}
H. Busack, physics.gen-ph/1006.3555.

\bibitem[\protect\citeauthoryear{Schiller {\it et al.} }{2012}]{Sch10}
S. Schiller {\it et al.}, \textit{Space Time Explorer and Quantum Equivalence Principle Space Test (STE-QUEST) Cosmic Vision 2015-2025 proposal}, (Proposal in response to ESA M3 call 2010); available at {\it http://www.exphy.uni-duesseldorf.de/Publikationen/2010/STE-QUEST\_final.pdf}.

\bibitem[\protect\citeauthoryear{Cacciapuoti {\it et al.} }{2012}]{SciRD}
L. Cacciapuoti {\it et al.}, \textit{STE-QUEST Science Requirements Document}, FPM-SA-DC-00001, issue 1, revision 2 (March 2012); available at {\it http://sci.esa.int/ste-quest/}.

\bibitem[\protect\citeauthoryear{Renk}{2012}]{Mag}
ESA/ESOC, \textit{STE-QUEST: Mission Analysis Guidelines (MAG)},  issue 2, revision 0 (June 2012).

\bibitem[\protect\citeauthoryear{Cacciapuoti {\it et al.} }{2007}]{ACES}L. Cacciapuoti {\it et al.}, \NP \textit{Proc. Suppl.} {\bf B 166}, 303 (2007).

\bibitem[\protect\citeauthoryear{Nobili {\it et al.} }{2007}]{GG}A. M. Nobili {\it et al.}, \CQG {\bf 29}, 184011 (2012).

\bibitem[\protect\citeauthoryear{Turyshev {\it et al.} }{2009}]{LATOR}S. G. Turyshev {\it et al.}, \textit{Exper. Astron.} {\bf 27}, 27 (2009).

\end{thebibliography}
\end{document}